\documentclass{article}

    \usepackage[preprint]{neurips_2024}

\usepackage[utf8]{inputenc} 
\usepackage[T1]{fontenc}    
\usepackage{hyperref}       
\usepackage{url}            
\usepackage{booktabs}       
\usepackage{amsfonts}       
\usepackage{nicefrac}       
\usepackage{microtype}      
\usepackage{xcolor}         
\usepackage{bm}
\usepackage{amsthm}
\usepackage{array}
\usepackage{color}
\usepackage{amsmath}

\usepackage{listings}
\usepackage{tikz}

\usepackage{graphicx}
\usepackage{caption}
\usepackage{subcaption} 

\title{Large Population Models}

\author{%
  Ayush Chopra \\
  Massachusetts Institute of Technology\\
  \texttt{ayushc@mit.edu} \\
}

\begin{document}

\maketitle

\begin{abstract}
Many of society's most pressing challenges—from pandemic response to supply chain disruptions to climate adaptation—emerge from the collective behavior of millions of individuals making decisions over time. Large Population Models (LPMs) offer an approach to understand these complex systems by simulating entire populations with realistic behaviors and interactions at unprecedented scale.

LPMs extend traditional modeling approaches through three key innovations: computational methods that efficiently simulate millions of individuals simultaneously, mathematical frameworks that learn from diverse real-world data streams, and privacy-preserving protocols that bridge simulated and physical environments. This allows researchers to observe how individual choices aggregate into system-level outcomes and test interventions before real-world implementation.

While current AI advances primarily focus on creating "digital humans" with sophisticated individual capabilities, LPMs develop "digital societies" where the richness of interactions reveals emergent phenomena. By bridging individual behavior and population-scale dynamics, LPMs offer a complementary path in AI research—illuminating collective intelligence and providing testing grounds for policies and social innovations before real-world deployment. We discuss the technical foundations and some open problems here. LPMs are implemented by the AgentTorch framework: \url{github.com/AgentTorch/AgentTorch}
\end{abstract}

\section{Introduction}
Many complex societal challenges - from pandemic risk to supply chain disruptions - emerge from the interactions of millions of individuals making decisions over time. Understanding and addressing these challenges requires computational tools that can: (1) simulate realistic populations at scale, (2) learn from heterogeneous data streams, and (3) integrate with real-world systems. 

Agent-based models (ABMs) offer a promising solution to capture these dynamics by simulating a collection of autonomous entities (called agents) that act and interact within a digital world. ABMs have proven useful across various domains, including epidemiology \cite{aylett2021june, kerrCovasimAgentbasedModel2021e, hinchOpenABMCovid19AgentbasedModel2021b}, economics \cite{axtell120MillionAgents2016, axtellAgentBasedModelingEconomics, carroHeterogeneousEffectsSpillovers2023}, and disaster response \cite{aylett2022epidemiological, ghaffarian2021agent}. Their bottom-up approach allows emergent phenomena to arise naturally from simple rules, offering insights into how individual behaviors aggregate to create system-level outcomes. However, their practical utility has been limited by three fundamental challenges:

\begin{itemize}
    \item \textbf{Scale}: ABMs struggle to scale to realistic population sizes while maintaining sophisticated agent behaviors. Recent work using large language models (LLMs) as agents has demonstrated more human-like decision-making~\cite{parkGenerativeAgentsInteractive2023, vezhnevetsGenerativeAgentbasedModeling2023a}, but is restricted to small populations of 25-1000 agents. Real-world policy decisions, however, require simulating millions of agents with both rich behaviors and complex interactions.
    \item \textbf{Data}: ABMs face significant challenges when assimilating heterogeneous data sources at scale. Current calibration approaches typically require computationally intensive sampling methods and struggle to incorporate streaming data or handle high-dimensional parameter spaces efficiently. The inability to seamlessly learn from diverse real-world data streams in real-time significantly constrains ABMs' predictive power.
    \item \textbf{Feedback}: ABMs typically treat agents as purely synthetic entities mimicking real-world counterparts. However, the increasing ubiquity of mobile and IoT devices creates opportunities for bidirectional feedback between simulated and physical agents. Realizing this potential requires frameworks that can handle both centralized simulation and privacy-preserving decentralized computation - a capability lacking in traditional ABM implementations.
\end{itemize}

\textbf{Contributions}:

\textbf{A.} This paper introduces Large Population Models (LPMs) as an evolution of ABMs that overcome these limitations through three key innovations: 
\begin{itemize}
    \item \textbf{Compositional Design}: LPMs enable efficient simulation of millions of agents on commodity hardware, overcoming traditional scale limitations through composable interactions and tensorized execution. This architectural innovation allows balancing behavioral complexity with computational constraints, enabling realistic agent behavior even at population scale.
    \item \textbf{Differentiable Specification}: LPMs make simulations end-to-end differentiable, supporting gradient-based learning for calibration, sensitivity analysis, and data assimilation. This enables efficient integration of heterogeneous data streams and composition with neural networks for improved prediction.
    \item \textbf{Decentralized Computation}: LPMs extend differentiable simulation and learning to distributed agents using secure multi-party protocols. This bridges the sim2real gap by allowing integration of real-world agents while protecting individual privacy.
\end{itemize}


\begin{figure}[h!]
    \centering
    \includegraphics[width=0.75\linewidth]{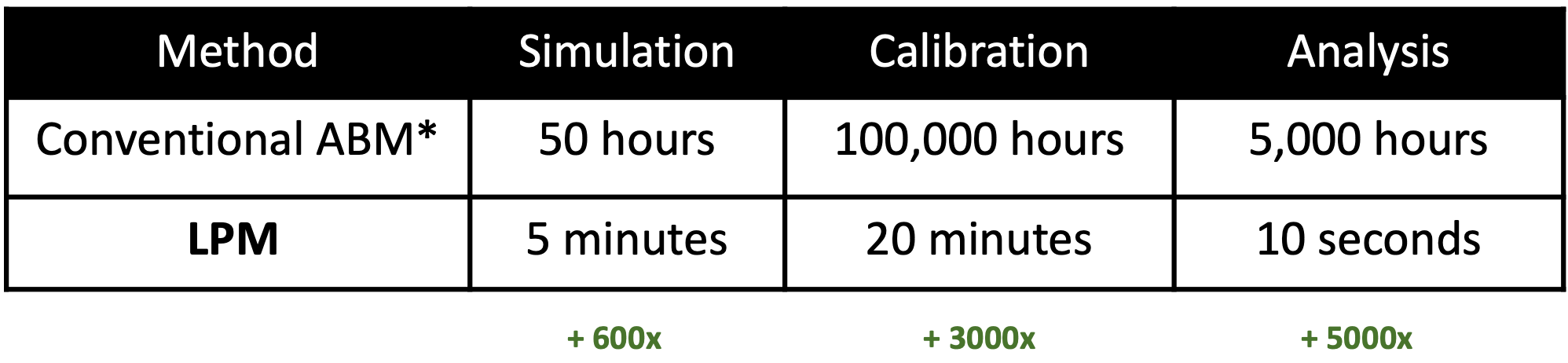}
    \caption{Performance benchmarking comparing computational efficiency of LPMs versus conventional ABMs for simulating 8.4 million agents representing NYC's population. LPMs demonstrate orders-of-magnitude improvements in simulation (600x), calibration (3000x), and analysis (5000x) runtimes, enabling previously infeasible large-scale agent-based modeling applications.}
    \label{fig:lpm_abm}
\end{figure}

\textbf{B.} We also introduce \textbf{AgentTorch}, an open-source framework that makes these theoretical advances of LPMs accessible and practical. AgentTorch uniquely integrates all critical capabilities needed for large-scale agent modeling : GPU acceleration, million-agent populations, differentiable environments, mechanistic modeling, LLM integration, and neural network composition. This allows LPM capabilities to work synergistically: scalable architecture enables efficient gradient computation, differentiability allows learning from distributed data, and decentralized protocols preserve these capabilities when integrating real systems. AgentTorch models are deploying LPMs around the globe - to help immunize millions of people by optimizing vaccine distribution strategies, and to track billions of dollars in global supply chains, improving efficiency and reducing waste.

Through detailed case studies on pandemic response in New York City, we demonstrate how these contributions enable more accurate predictions, more efficient policy evaluation, and more seamless integration with real-world systems than traditional ABMs. This evolution in modeling capability opens new possibilities for addressing complex societal challenges that emerge from individual behaviors and interactions.
\section{Preliminaries}
~\label{sec:prelim}

Agent-based models (ABMs) offer a bottom-up approach to simulating complex systems by representing how individual entities ("agents") act and interact within computational environments. Unlike equation-based models that describe aggregate behaviors, ABMs capture the heterogeneity of individual actions and local interactions, allowing emergent phenomena to arise naturally from simple rules. To clarify terminology: in ABMs, "agents" refer to computational entities that represent real-world individuals with autonomous behaviors, distinct from "AI agents" that execute tasks on behalf of users. ABM agents are designed to mimic real-world behaviors through rule-based decision processes, executed over discrete timesteps, ultimately revealing how individual decisions aggregate to population-level outcomes.

\subsection{Formal Representation}
Consider a population of $N$ individuals. We denote by $\bm s_i(t)$ the state of individual $i$ at simulation time $t$ which contains both static and time evolving properties of individuals. For instance, $\bm s$ may represent the age and disease status of humans in epidemiological models, or the account balance of firms in a financial auction model. 

As the simulation proceeds, an individual $i$ updates their state $\bm s_i(t)$ by interacting with their neighbors $\mathcal N_i(t)$ and their environment $e(t)$, which can both also be time evolving. The neighborhood of an individual can be specified using a graph, a proximity method or other methods. 

We denote by $m_{ij}(t) = M(\bm s_j(t), e_{ij}(t), \boldsymbol{\theta}, t)$ the message or information that individual $i$ obtains from their interaction with neighbor $j$. In an epidemiological model, this may represent the transmission of infection from individual $j$ to individual $i$, dependent upon the infectivity of $j$ ($\bm s_j$), the properties of the virus ($\boldsymbol{\theta}$) and the nature of the interaction ($e_{ij}$). 

The individual's outcome also depends upon their behavior, modeled as $\ell(\cdot | \bm s_i(t))$. This behavior function represents the decisions an individual makes given their state and environmental context—for example, whether they choose to isolate, wear masks, or participate in social gatherings. Thus, at step $t$, each agent updates its state as:
\begin{equation}
    \label{eq:agent_update}
    \bm s_i(t+1) =  f\left(\bm s_i(t), \bigoplus_{j \in \bm N_i(t)} m_{ij}(t), \; \ell(\cdot | \bm s_i(t)), \; e(t; \boldsymbol{\theta})\right)
\end{equation}
where $\oplus$ denotes an aggregation over all received messages. Similarly, the environment can also have it's own dynamics that depend upon the agents' updates and actions,
\begin{equation}
    \label{eq:env_update}
    \bm e(t+1) = g\left(\bm s(t), e(t), \bm \theta  \right).
\end{equation}
For instance, in an epidemiological model, $e$ captures the dynamics of disease transmission, viral evolution, vaccination protocols, etc.

The specific choices of $f$ and $g$ define the dynamics of the simulation and they are typically stochastic functions which can be mechanistically specified or learned from data.

\subsection{Core Tasks}
ABMs enable three tasks that form the basis for understanding and analyzing complex systems:

\textbf{Simulation} involves initializing individual and environment states $(\bm{s}(0), e(0))$ and recursively applying the update rules in Equations \ref{eq:agent_update} and \ref{eq:env_update} over multiple timesteps. While the state space of the simulation is vast, we typically focus on aggregate outcomes represented as time series $\bm{x}_t = h(\bm{s}(t))$. For example, epidemiological simulations might track the daily count of infected individuals across different demographic groups, while economic simulations might monitor unemployment rates and market indicators. Formally, the simulation can be expressed as a stochastic map:
\begin{equation}
    \label{eq:sim_eqn}
    \bm x = F(\bm \theta, \bm s(0), \bm e(0)),
\end{equation}
where $F = (f, g) \circ \dots \circ (f, g)$ represents the composition of agent and environment update functions repeated for $T$ timesteps. The computational complexity of this process grows with both population size and behavioral sophistication, creating a tension in traditional ABM implementations.

\textbf{Calibration} aims to identify parameters $\hat{\bm{\theta}}$ (or a distribution over parameters) that make simulation outputs consistent with observed data. Formally, calibration refers to the process of tuning structural parameters $\boldsymbol{\theta}$ so that simulation outputs $\mathbf{x}$ are compatible with given observational data $\mathbf{y}$. In epidemiological models, for instance, this entails determining values for parameters like the reproduction number and mortality rates to align with observed infection or mortality data.

Due to the stochasticity of ABMs and their partial observability, multiple sets of parameter values $\boldsymbol{\theta}$ may be compatible with the observed data $\mathbf{y}$. Consequently, accurate uncertainty estimation around calibrated parameters becomes essential. Incorporating expert knowledge (e.g. external data signals) that may indicate preferences for certain regions of the parameter space is also important for robust calibration.

These requirements can be addressed through a Bayesian framework, where parameter inference corresponds to determining the posterior distribution over parameters, $\pi(\boldsymbol{\theta} \mid \mathbf{y})$ using Bayes' theorem:
$\pi(\boldsymbol{\theta} \mid \mathbf{y}) = \frac{p(\mathbf{y}\mid \boldsymbol{\theta}) ;\pi(\boldsymbol{\theta})}{p(\mathbf{y})}$,
where $\pi(\boldsymbol{\theta})$ is the prior distribution, $p(\mathbf{y}\mid \boldsymbol{\theta})$ is the likelihood function, and $p(\mathbf{y})$ is the marginal likelihood. For ABMs, the likelihood function is typically intractable, necessitating likelihood-free calibration algorithms.


In practice, calibration encompasses a spectrum of approaches, from manual parameter tuning guided by domain expertise~\cite{romero2021public} to sophisticated Bayesian techniques that quantify uncertainty~\cite{hinchOpenABMCovid19AgentbasedModel2021b,aylett-bullockJuneOpensourceIndividualbased2021b}. The optimization becomes challenging with computational complexity and stochasticity of $F$ and the high dimensionality of $\bm{\theta}$.


\textbf{Analysis} leverages calibrated models to understand system dynamics, explore counterfactuals, and inform decision-making. What distinguishes ABMs in analysis tasks is their ability to trace macroscopic outcomes to specific agent behaviors and interactions, providing mechanistic insights that purely statistical models cannot. 

Retrospective analysis examines how observed outcomes emerged from interactions. For example,~\cite{martinSociodemographicHeterogeneityPrevalence2020a} analyzed how sociodemographic factors contributed to COVID-19 disparities identifying how risk varies across demographics. Counterfactual analysis explores "what-if" scenarios by varying conditions or interventions.~\cite{romero2021public} evaluated delaying second vaccine doses to identify conditions where such delays could reduce overall mortality—directly informing vaccination policies in multiple countries. These capabilities address the Lucas critique—the observation that policy effects cannot be predicted solely from historical data when agents can adapt their behavior in response to policies~\cite{LUCAS197619}. By explicitly modeling behavioral adaptation through $\ell(\cdot | \bm{s}_i(t))$, ABMs capture how interventions reshape the decision rules that generate outcomes, providing more robust guidance for policy design.

\section{Case Study: COVID-19 in New York City}
Imagine you are the public health leader for New York City during COVID-19. You want to understand how the disease is spreading and implement interventions to stop it. Consider some pertinent questions: When will the next wave emerge? Which test is better: PCR vs AntiGen? What if we give a \$500 stimulus check? The answers to these questions are shaped by the interplay of citizen behavior, transmission dynamics and intervention design. By simulating their complex interactions over a large population, we can understand feedback loops and then optimize strategies to control the spread of disease. For instance, we can understand how individual behavior change (e.g. fatigue) can seed a pandemic; and why prioritizing test speed over accuracy can be better for an individual, even if it may not seem like it. Large Population Models provide a sandbox to design and evaluate such policies, by simulating millions of autonomous agents at population-scale.


\subsection{Population and Environment Definition}
We construct a synthetic population of $N = 8.4$ million individuals representing New York City, with demographic profiles derived from the 2022 American Community Survey. Each agent state $\bm{s}_i(t)$ includes static attributes (age, gender, income, occupation) derived from census and dynamic properties (disease status, employment status) that evolve through the simulation. Interactions occur over household, workplace, and mobility networks, with recreational and workplace mobility parameterized using Google Mobility trends~\cite{santana2023covid}.

The environment $e(t)$ captures two interconnected systems: a contact-based disease model where infection spreads through agent interactions, and a labor market model tracking employment and financial conditions. Agents interact through networks $\mathcal{N}_i(t)$ constructed from household, workplace, and mobility data, which serve as channels for both disease transmission and economic influence.





\subsection{Policy Question: Impact of Stimulus Payments}
We focus on a specific policy question: "What if we give stimulus checks?". This intervention generated complex effects during the pandemic through behavioral adaptation~\cite{li2021impacts}. Stimulus programs introduced to mitigate economic hardship and encourage compliance with health measures had unintended effects on labor markets and resource allocation~\cite{falcettoni2020acts}. This exemplifies the complex feedback loops between individual behavior adaptation $\ell(\cdot|\bm{s}_i(t))$, disease transmission through agent interaction networks, and economic conditions affected by aggregate behavioral choices. 

\begin{figure}[h!]
    \centering
    \includegraphics[width=0.45\linewidth]{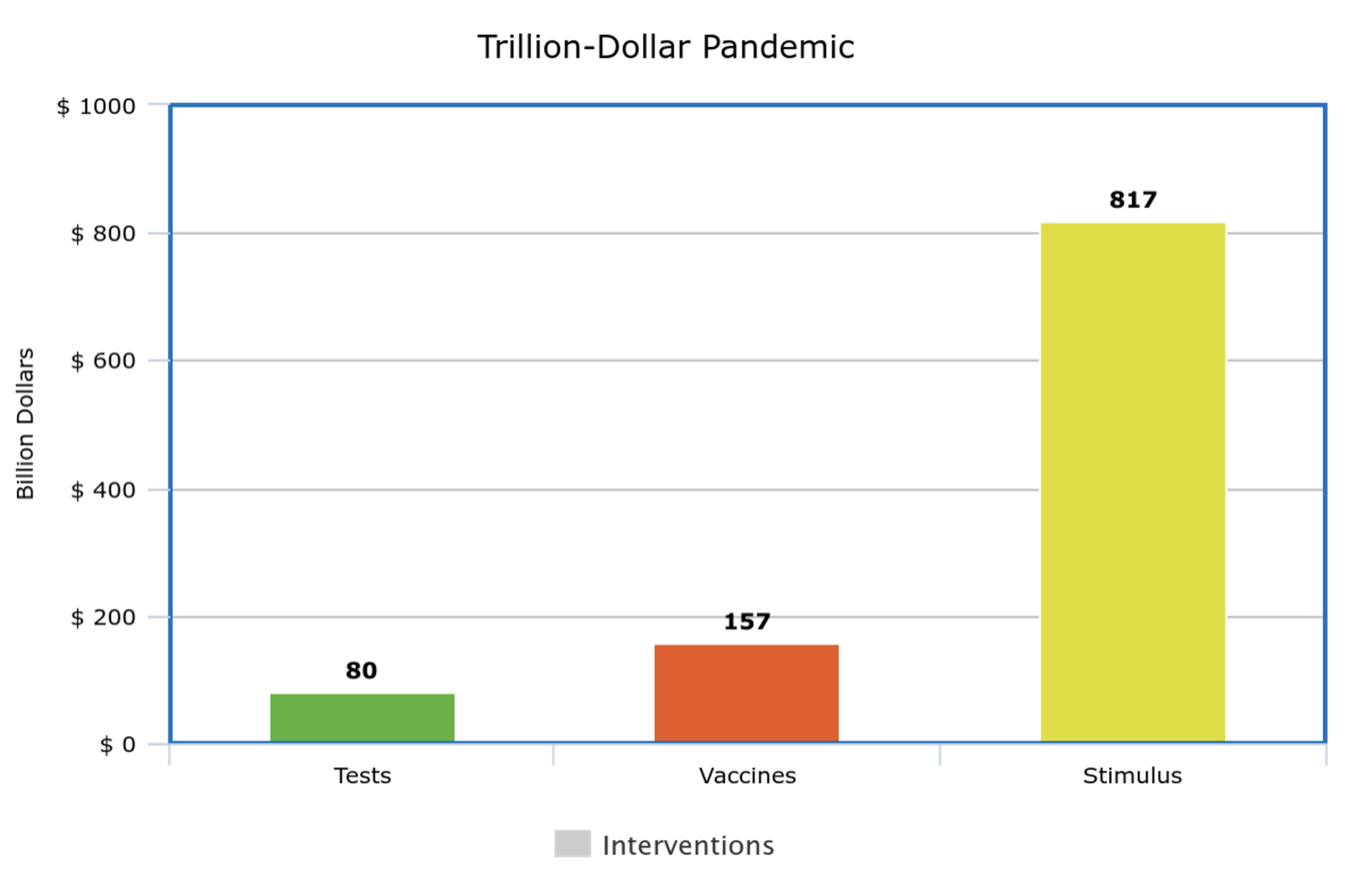}
    \caption{Trillion-Dollar Pandemic resource allocation breakdown demonstrating the critical role of behavioral incentives in pandemic management. While \$200B was allocated to infection control measures (tests and vaccines), \$800B was directed toward stimulus payments to influence citizen behavior, highlighting the importance of modeling behavioral feedback loops in pandemic dynamics.}
    \label{fig:covid-expense}
\end{figure}

\subsection{Fundamental Challenges in Population Modeling}
Modeling these dynamics for NYC reveals three fundamental challenges that limit traditional ABMs:

\textbf{Challenge 1: Scale vs. Expressiveness Trade-off}:  Simulating nuanced behavioral responses for 8.4 million individuals across multiple networks (household, workplace, transit) quickly exceeds practical computational constraints using conventional approaches. Traditional epidemiological models can simulate realistic population sizes but rely on simplified behavioral rules for agent and environment that fail to capture the nuanced decision-making of individuals during a pandemic. Conversely, recent LLM-based approaches demonstrate sophisticated adaptive behaviors but remain limited to small populations of 25-1000 agents and capture unrealistic environments. This creates a fundamental tension: sophisticated behavior OR population scale, but not both.

\textbf{Challenge 2: Heterogeneous Data Integration}: NYC health officials received diverse data streams—clinical reports from hospitals, mobility patterns from cell phones, economic indicators from government agencies, and survey data on compliance behaviors—each with different granularity, reliability, and privacy constraints. Traditional calibration approaches struggle to efficiently incorporate these diverse signals. Real-world data often comes with diverse uncertainties, reporting frequencies, and granularities, making integration challenging without a unified framework. Privacy constraints further complicate calibration, as critical data may be inaccessible due to regulations.

\textbf{Challenge 3: Simulation-Reality Gap}: Data privacy and availability constraints limit the quality of simulations which provides limited visibility into individual behavioral. "Pandemic fatigue" reduced compliance with health measures over time, and financial pressures from depleted savings altered risk calculations despite unchanged disease threat. These complex behavioral adaptations created a disconnect between model predictions and observed outcomes. Critical parameters in traditional models often conflate behavioral and environmental factors, making it difficult to isolate the effects of interventions. The lack of bidirectional feedback between simulated and real-world agents further exacerbates this problem.

This pandemic scenario provides an ideal testbed to demonstrate capabilities because it requires understanding complex feedback loops between behavioral adaptation, disease dynamics, and economic impacts at population scale—precisely the challenges that LPMs are designed to address.


The pandemic scenario provides an ideal testbed to demonstrate capabilities because it requires understanding complex feedback loops between behavioral adaptation, disease dynamics, and economic impacts at population scale—precisely the challenges that LPMs are designed to address.

\begin{figure}[t!]
    \centering
    \includegraphics[width=0.85\linewidth]{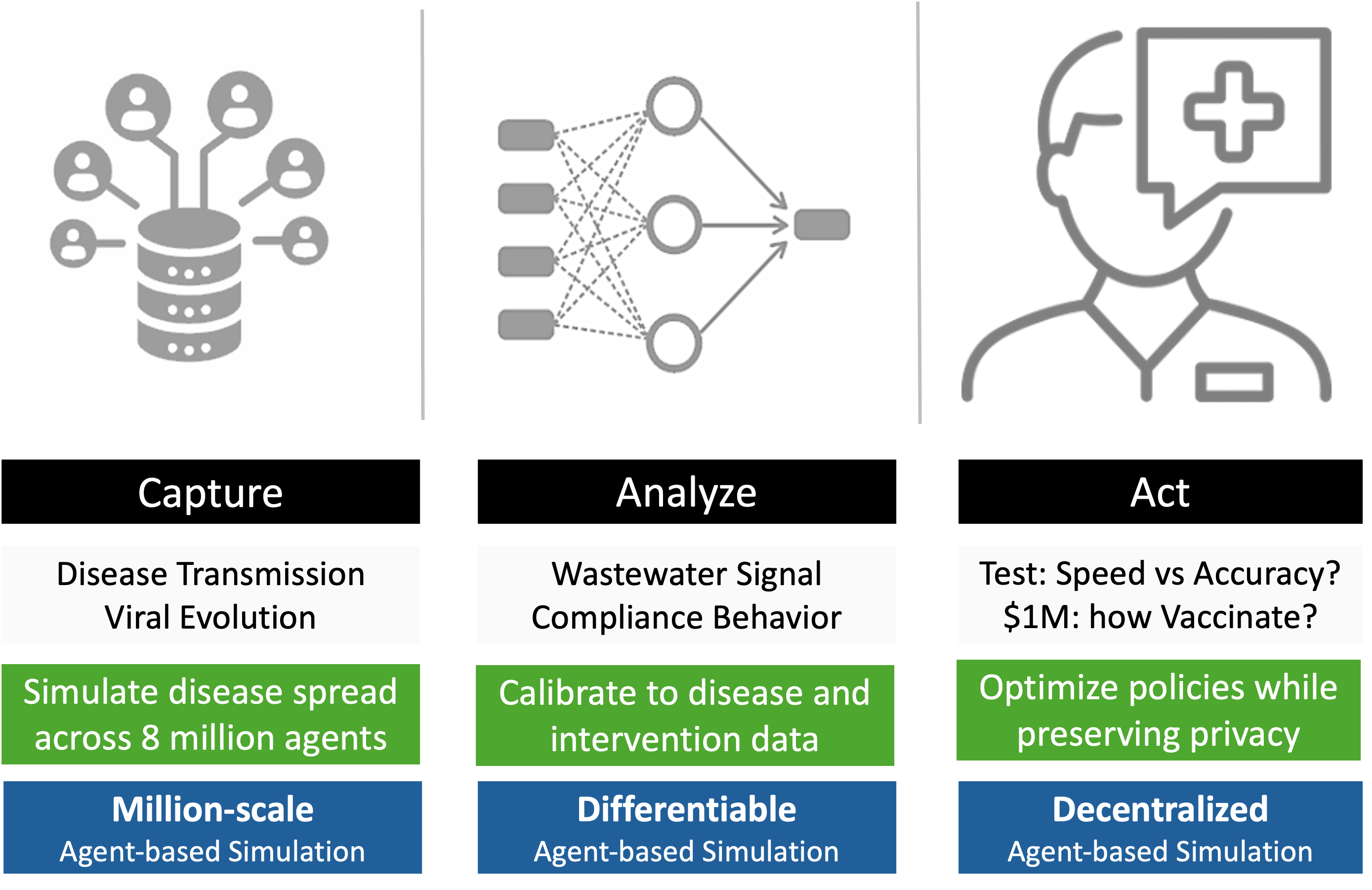}
    \caption{Research Pillars: LPMs alleviate the three challenges of ABMs by making simulations scalable, differentiable and decentralized. In Public Health, this enables LPMs to capture disease transmission and viral evolution across millions of agents; analyze outcomes by ingesting multi-modal disease, behavior and intervention data and securely deploy policies while preserving privacy.}
    \label{fig:lpm_tech}
\end{figure}



\section{Challenge 1: Scale vs Expressiveness Trade-off}
Consider a day in the life of a New York City resident during COVID-19: they leave their apartment, commute on crowded subway cars, and interact with coworkers. Throughout the day, they make numerous decisions based on their circumstances—whether to wear a mask, get tested after possible exposure, or reduce social activities due to financial constraints. Their behavior evolves based on test results, symptoms, financial pressures, and social influences.

To understand pandemic dynamics, we need to model 8.4 million such individuals and their interactions. This creates an immense computational challenge: representing realistic behavior for millions of agents interacting across multiple networks quickly becomes intractable with traditional approaches. Conventional epidemiological models like those referenced in ~\cite{romero2021public,aylett-bullockJuneOpensourceIndividualbased2021b} can simulate realistic population sizes but rely on simplified behavioral rules that fail to capture nuanced decision-making during a pandemic and are not compatible with data-driven learning. Conversely, recent LLM-based approaches~\cite{parkGenerativeAgentsInteractive2023} demonstrate sophisticated adaptive behaviors but remain limited to small populations of 25-1000 agents and unrealistic environments - far from the scale needed to model metropolitan dynamics.

This creates a fundamental tension: sophisticated behavior OR simulation scale, but not both. Yet the critical insights emerge precisely from their interaction—how individual adaptive behaviors aggregate to population-level outcomes, and how those outcomes in turn influence individual decisions.

Further, traditional ABM approaches often force artificial boundaries between these systems—epidemiologists focus on disease dynamics while economists model financial interventions, each using different mathematical formalisms. Yet the critical insights emerge precisely from their interactions: How does a \$600 stimulus check affect isolation behavior among different demographic groups? How might changed mobility patterns accelerate viral evolution? These questions require a unified computational representation that can seamlessly compose multiple systems.

\textbf{Key Insight}: We resolve this tension through a composable domain-specific language (FLAME) and an agent archetype-based approach. FLAME decomposes complex environmental dynamics into modular, composable substeps that can be efficiently executed through tensorized operations. For agent behavior, our breakthrough insight is that while agent states are highly heterogeneous, their decision-making processes often follow similar patterns based on demographic, socioeconomic, and behavioral characteristics. This allows us to identify representative agent archetypes that capture the essence of behavioral variations while dramatically reducing computational requirements. This is represented in the following papers:
\begin{itemize}
    \item Chopra et al. flame: a framework for learning in agent-based models (AAMAS 2024, Oral) \href{https://dl.acm.org/doi/abs/10.5555/3635637.3662888}{[link here]}
    \item Chopra et al. on the limits of agency in agent-based models (AAMAS 2025, Oral) \href{https://arxiv.org/pdf/2409.10568}{[link here]}
    \item Romero-brufau, Chopra et al. Public health impact of delaying second dose of BNT162b2 or mRNA-1273 covid-19 vaccine (British Medical Journal 2021) \href{https://www.bmj.com/content/373/bmj.n1087}{[link here]}
\end{itemize}

\subsection{Contribution 1: Composable Domain-specific Language}
A comprehensive simulation must capture multiple interacting systems across different spatial and temporal scales. For instance, an environment of pandemic dynamics, $e$, must simultaneously represent disease transmission events across contact networks (discrete-time models), individual viral evolution (continuous-time with differential equations), mobility patterns (learned via neural networks) and stimulus interventions (discrete-event logic). These diverse dynamics must be coordinated across individual, household, neighborhood, and city-level scales, at each step of the simulation.

We introduce FLAME \cite{chopra2024framework}, a domain-specific language that enables flexible specification of complex environmental dynamics through composable interactions. This framework extends the environment update function $e(t + 1) = g(s(t), e(t), \theta)$ defined in Section~\ref{sec:prelim} to support multiple modeling paradigms and interaction scales. FLAME addresses this challenge through three key capabilities: modular substep definition, tensorized execution, and gradient-based composition.

\begin{figure*}[h!]
    \centering
    \includegraphics[width=0.95\textwidth]{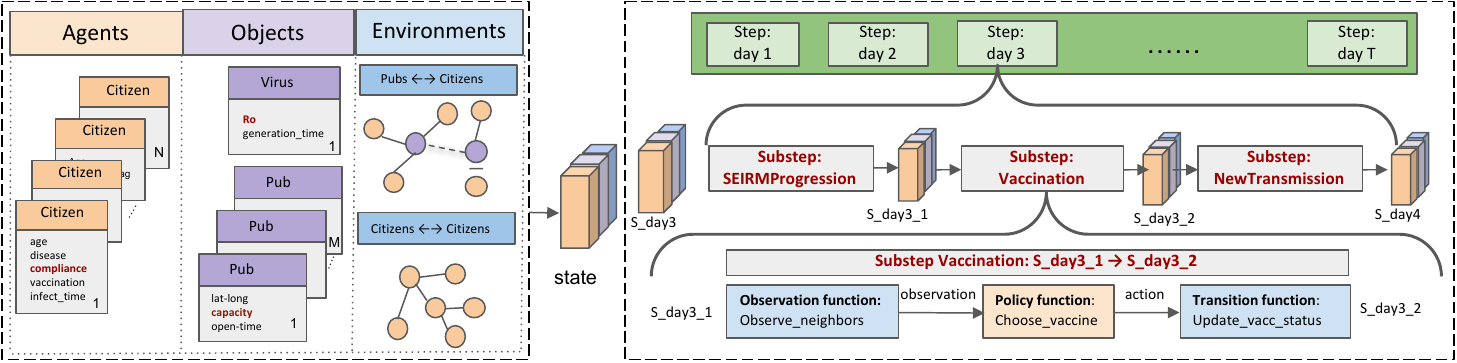}
    \caption{\textbf{Domain Specific Language}. The simulation has N citizens (Agents) that interact directly and co-locate across pubs (Object) to spread the virus (Object). The simulator state is a collection of properties that describe  these entities, is initialized once, and transformed during T simulation steps. Each step models the disease progression of infected agents ({SEIRMProgression}), vaccination of susceptible agents ({Vaccination}), and transmission of new infections ({NewTransmission}) to recursively transform the simulation state over these substeps. FLAME ensures gradient flow through all simulation steps and enables automatic differentiation of any state property or substep function.}
    \label{fig:framework}
\end{figure*}

\textbf{First, functional architecture}: FLAME decomposes environment dynamics into modular substeps, each represented as a functional mapping $\Phi_i: S \rightarrow S$ that transforms the simulation state. Each substep follows an observation-policy-transition pattern, where observation functions $o: S \rightarrow O$ extract relevant state features, policy functions $\pi: H \rightarrow A$ determine actions based on observation histories, and transition functions $t: S \times A \rightarrow S$ update states accordingly. This decomposition enables precise modeling of heterogeneous dynamics—a substep might represent virus transmission events through a mechanistic contact model, while another might capture mobility patterns through a learned neural network.

\textbf{Second, tensorized execution}: FLAME implements tensorized execution to simulate these substeps across billions of potential interactions among millions of agents. For large populations, direct simulation of all pairwise interactions quickly becomes intractable. FLAME transforms these interactions into sparse tensor operations that can be efficiently executed on commodity GPU hardware. This optimization relies on two principles: (1) \underline{small-world interaction networks}, where most agents interact with a limited subset of the population, enabling sparse tensor representations; and (2) \underline{permutation invariance within substeps}, where the order of operations \textit{in a simulation step} doesn't affect the outcome. For instance, when modeling disease transmission across households on a given day, the likelihood of infection for an agent depends on their total exposure, not the sequence of individual contacts.

\textbf{Third, differentiable design}: FLAME ensures modularity of substeps via differentiable specification. Formally, each environment substep $\Phi_i$ is differentiable if, given a smooth objective $L = f(\Phi_i(S_T))$ defined on the simulation state, the gradient $\nabla_\theta L$ exists and can be computed. Rather than approximating gradients or replacing stochastic components with surrogate models, FLAME maintains differentiability by applying reparameterization techniques to stochastic mechanisms. For example, discrete contact events are reparameterized to preserve gradient flow while maintaining realistic stochasticity.This differentiability of each substep enables gradient-based composition, via the chain rule, and allows modular design of complex simulations. As we see later, this approach enables efficient calibration, sensitivity analysis, and optimization across scales—allowing researchers to investigate complex feedback loops like how viral properties might adapt in response to city-level case statistics.

These capabilities dramatically transform our ability to model complex environments. By decomposing intricate dynamics into composable, differentiable substeps and executing them efficiently at scale, FLAME achieves a 200× speedup compared to traditional implementations when simulating 8.4 million agents \cite{quera2023don}. This makes previously intractable scenarios computationally feasible, enabling more accurate representation of the environments within which agents interact.

\begin{figure}
    \centering
    \begin{subfigure}[b]{0.45\textwidth}
        \centering
        \includegraphics[width=\textwidth]{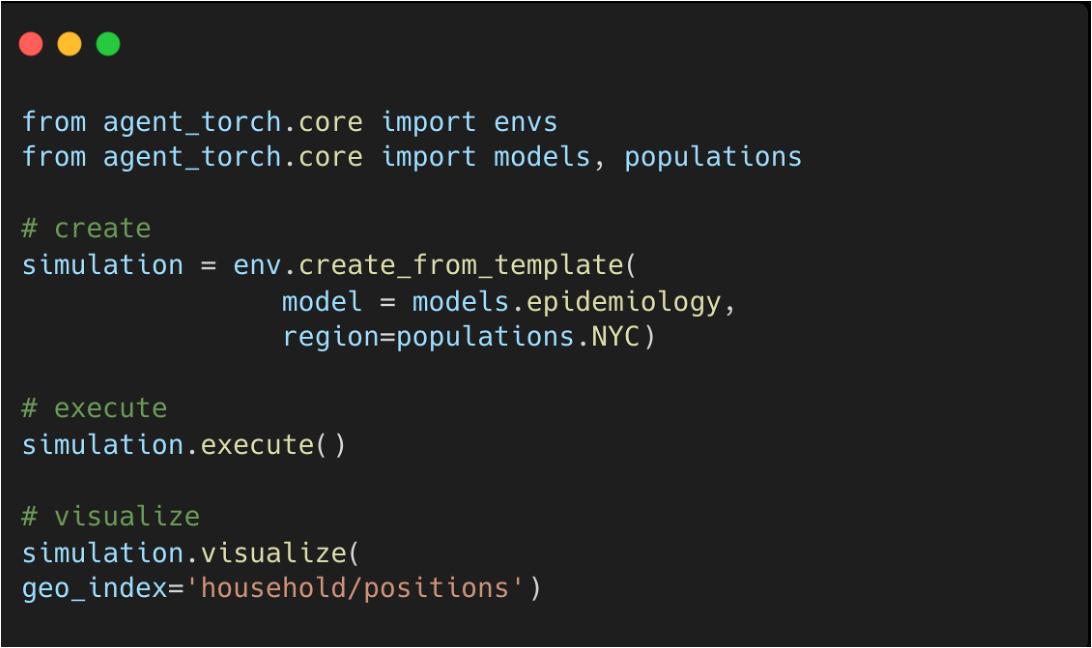}
        \caption{}
        \label{fig:flame_1}
    \end{subfigure}
    \hfill
    \begin{subfigure}[b]{0.45\textwidth}
        \centering
        \includegraphics[width=\textwidth]{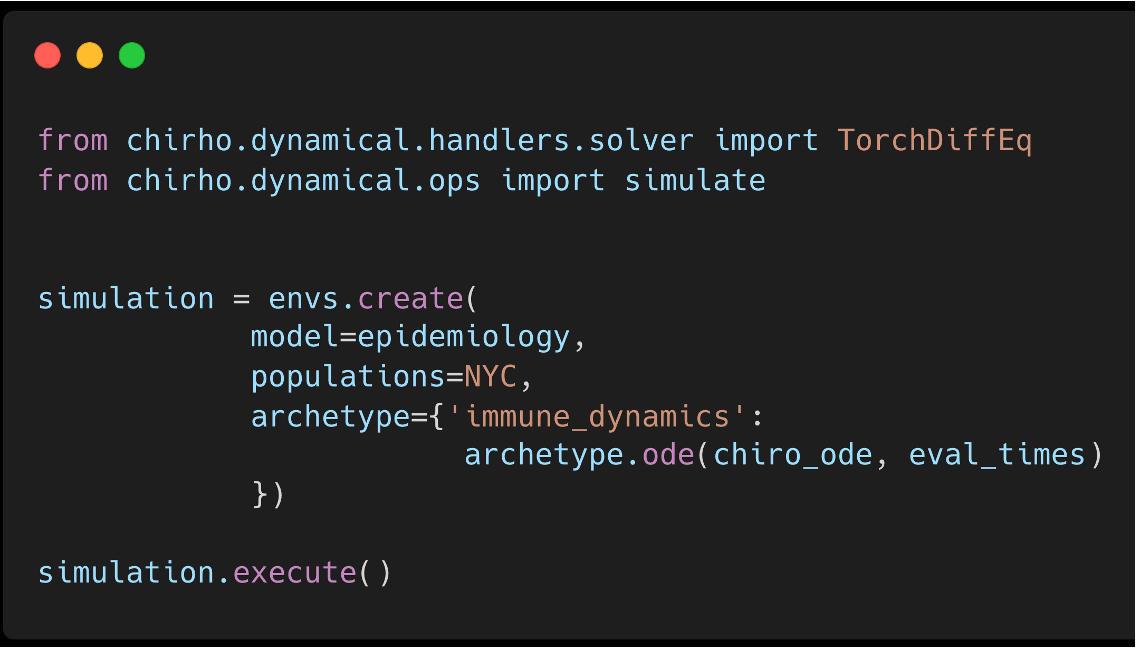}
        \caption{}
        \label{fig:flame_2}
    \end{subfigure}
    \caption{Multi-fidelity simulations. (a) Implementation of the FLAME architecture within the AgentTorch API, with code snippet for creating, executing, and visualizing simulations with 8.4 millions NYC agents. (b) AgentTorch API demonstrating integration of continuous-time ODE models for immune dynamics with the discrete-time disease simulation.}
\end{figure}

\subsection{Contribution 2: Scaling LLM-guided Behavior for Million Agents}
Having established a framework to efficiently model complex environments, we face the challenge of simulating realistic agent behaviors within these environments. From Section~\ref{sec:prelim}, an agent's state update depends on both environmental factors and behavioral decisions, where $\ell(\cdot|s_i(t))$ models the agent's behavioral function.

In a realistic simulation, agents must make numerous context-dependent decisions throughout each time step—whether to wear masks, get tested after potential exposure, reduce social activities due to financial constraints, or seek vaccination. These behaviors evolve dynamically based on extrinsic interventions - test results, lockdown mandates, and social influences from their network or even intrinsic adaptations - fatigue, symptom progression, financial pressures. Capturing this behavioral complexity is essential for realistic outcomes, particularly when behavioral adaptation creates feedback loops with environmental dynamics.

However, implementing sophisticated agent behavior at scale presents a fundamental computational challenge. Recent approaches using Large Language Models (LLMs) \cite{parkGenerativeAgentsInteractive2023} have demonstrated impressive capabilities in generating context-aware, adaptive agent behaviors, but remain limited to small populations of 25-1000 agents. First, naively applying such approaches to millions of agents would require billions of model queries per simulation step—computationally infeasible and economically prohibitive. Second, our analysis shows the 'limits of agency' highlighting the crucial of population-scale in shaping real-world outcomes~\cite{chopra2025limits}.

We resolve this dilemma through a novel agent archetype-based approach that dramatically reduces computational requirements while preserving behavioral sophistication. Our key insight is that while agent states are highly heterogeneous, their individual decision-making processes often follow similar patterns based on demographic, socioeconomic, and behavioral characteristics. For instance, we can jointly prompt agent in age group 31-40 instead of distinctly identify age as 31 vs 34. This assumption is also practical since census data is often aggregated at demographic-level resolution (giving age-group distribution, instead of age distributions, for cohorts). We formalize this insight through the following methodology:

First, we identify representative agent archetypes based on key factors that influence decision-making. For pandemic behaviors, these might include static attributes - age groups, income brackets, occupation categories, and baseline compliance tendencies, or dynamic properties - disease status, employment status, family obligations. We note two points here: i) the number of archetypes may dynamically evolve throughout the simulation; ii) an agent may change it's archetype associations throughout the simulation. For a large population, the number of archetypes, at any time $t$, $K_t$ is typically several orders of magnitude smaller than the population size $N$.

Second, for each archetype $k$ ($\in K_t$) and action $\alpha$ (e.g., "decide to isolate"), we estimate the probability distribution over possible decisions using:
\begin{align}
p_{\alpha}(k, t) &= \mathbb{E}[\ell(\cdot|s_k(t), e(t), \theta)] \\
&\approx \frac{1}{M} \sum_{j=1}^{M} \xi_j \text{ with } \xi_j \sim \ell(\cdot|s_k(t), e(t), \theta)
\end{align}

where $\ell$ represents an LLM-based decision function queried $M$ times to account for response variability. By prompting representative archetypes instead of each unique individual agency, we capture a distribution of how agents in the archetype respond to the current environmental context.  

Third, each individual agent $i$ samples a specific action based on the archetype $k$ it associates with:

\begin{align}
\alpha_i(t) \sim \text{Categorical}(p_{\alpha}(k, t))
\end{align}

This approach enables modeling nuanced behavioral phenomena at metropolitan scale with just $K \times A \times M$ queries (for $K$ archetypes, $A$ actions, and $M$ samples per distribution) instead of $N$ queries (for $N$ agents). For large-scale simulations where $N \gg K$, this offers computational savings of several orders of magnitude without sacrificing behavioral richness.

\begin{figure}
    \centering
    \begin{subfigure}[b]{0.55\textwidth}
        \centering
        \includegraphics[width=\textwidth]{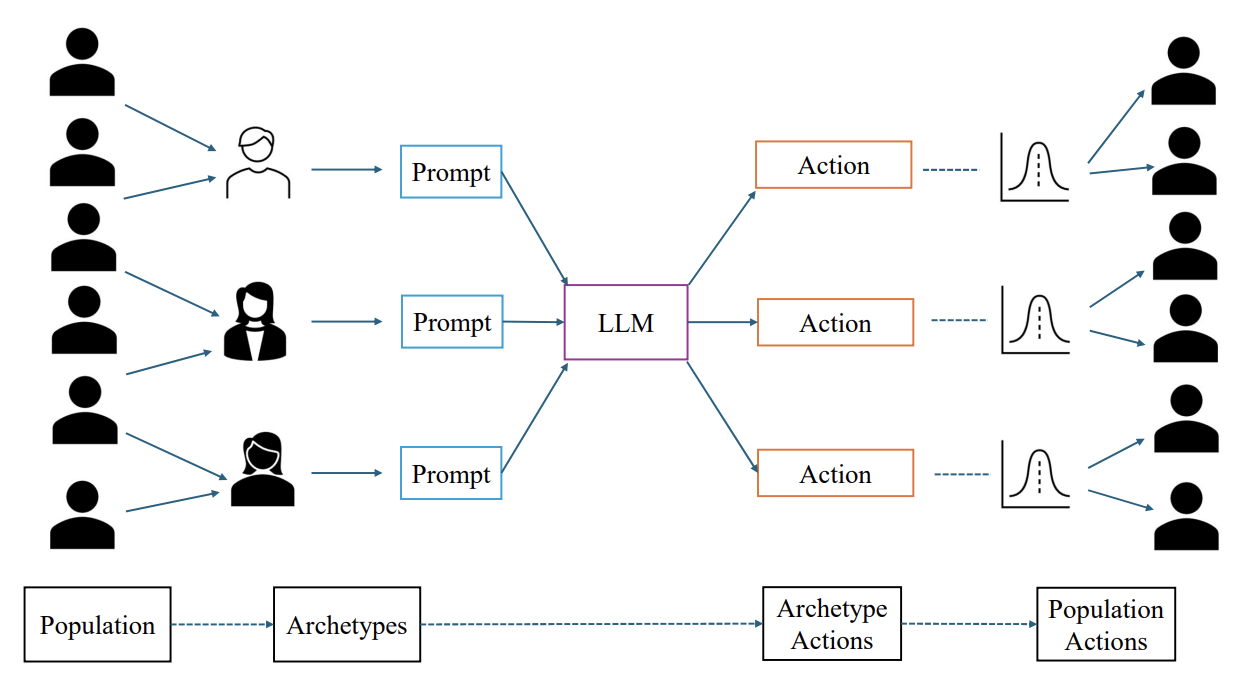}
        \caption{}
        \label{fig:arch_1}
    \end{subfigure}
    \hfill
    \begin{subfigure}[b]{0.43\textwidth}
        \centering
        \includegraphics[width=\textwidth]{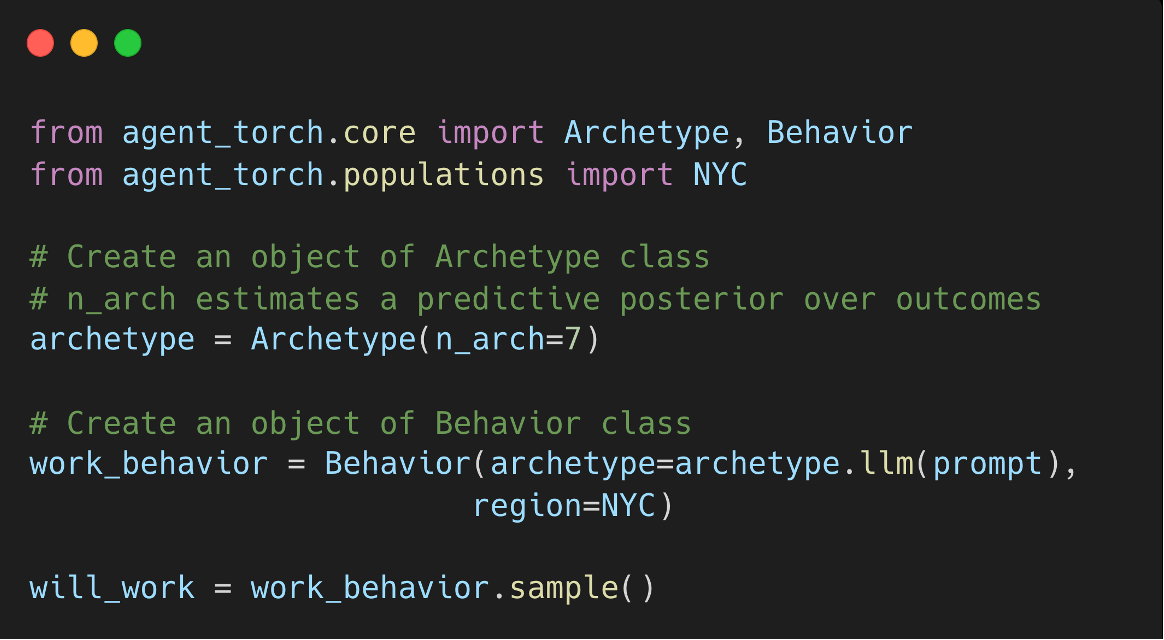}
        \label{fig:arch_2}
        \caption{}
    \end{subfigure}
    \caption{Modeling behavior for 8.4 million agents over 90 steps for \$500 (a) LPMs model distinct behavior for each agent in the simulation, at each simulation step. To reduce computational overhead, we utilize archetype-based prompting where we prompt representative agent archetypes instead of individual agents and individual agents sample their behavior from corresponding archetypes. This allows to balance individual behavior with population-scale which improves simulation performance compared to either prioritizing population-scale or individual agency. (b) AgentTorch Archetype API code sample showing implementation of the behavior estimation framework for efficient large-scale behavior modeling.}
\end{figure}
Importantly, this approach preserves both inter-group and intra-group heterogeneity. Even agents belonging to the same archetype may make different decisions due to the probabilistic sampling, while archetypes themselves capture major demographic and behavioral differences. Additionally, an agent's archetype assignment may change over time in response to evolving circumstances, enabling the simulation to capture phenomena such as:
\begin{itemize}
\item Time-dependent adaptation ("pandemic fatigue"), where compliance with health measures decreases over extended periods
\item Context-sensitive responses to financial incentives that vary by socioeconomic status
\item Social influence effects, where behavioral norms propagate through community networks
\end{itemize}

By maintaining heterogeneity through probabilistic sampling while drastically reducing computational requirements, archetype-based behavior modeling makes sophisticated agent behaviors feasible at population scale. This capability enables exploration of realistic behavioral interventions—a critical requirement for designing effective public health policies.

\section{Challenge 2: Heterogeneous Data Integration}
To use these simulations effectively, we need to calibrate them to real-world data. During the pandemic, decision-makers faced a paradoxical challenge: data abundance coupled with information scarcity. Public health departments received heterogeneous streams of information—clinical reports, wastewater sampling, mobility patterns, social media sentiment—yet struggled to integrate these signals into coherent, actionable insights. 

This challenge manifested in several ways:
\begin{itemize}
    \item First, each data source provided only a partial, noisy glimpse into the underlying dynamics. Case counts were subject to testing availability and reporting delays; mobility data captured movement but not purpose and were often a lagging indicator; economic indicators reflected aggregate impacts but not causal mechanisms.
    \item Second, complex interdependencies between signals made interpretation difficult. For example, declining mobility might indicate either successful isolation policies or economic hardship forcing essential workers to change schedules. Distinguishing between these scenarios requires simultaneous integration of multiple data streams.
    \item Third, the stochastic nature of both the disease process and human behavior introduces substantial uncertainty. Traditional calibration approaches require prohibitively large numbers of simulation runs, particularly for models with millions of agents and hundreds of parameters.
\end{itemize}

Traditional ABMs rely on offline calibration algorithms to train surrogate models on trajectories sampled from the simulator using blackbox inference techniques like Approximate Bayesian Computation~\cite{dyerBlackboxBayesianInference2024}. These techniques are sample inefficient (requiring large number of simulated trajectories) and don't scale to high-dimensional parameter spaces.

\textbf{Key Insight}: The power of LPMs emerges from the differentiability of our simulation framework. Since each protocol in FLAME is differentiable, by the chain rule of calculus, their composition maintains differentiability throughout the entire simulation. This transforms LPMs from black-box simulators into first-class computational objects amenable to modern optimization techniques. By applying automatic differentiation, we can efficiently compute gradients of simulation outputs ($\bm x$) with respect to parameters $\nabla_{\boldsymbol{\theta}}(x)$, enabling gradient-based calibration and zero-shot sensitivity analysis without requiring surrogate models or prohibitively large numbers of simulation runs.

This is demonstrated in the following publications:
\begin{itemize}
    \item Chopra et al. differentiable agent-based epidemiology (AAMAS 2023, Oral) \href{https://dl.acm.org/doi/abs/10.5555/3635637.3662888}{[link here]}.
    \item Quera-bofarull, Chopra et al. Don't simulate twice: one-shot sensitivity analysis via automatic differentiation (AAMAS 2023, Oral) and (ICML-W 2022 Best Paper Award) \href{https://www.ifaamas.org/Proceedings/aamas2023/pdfs/p1867.pdf}{[link here]}
    \item Garg and Chopra. Distributed Calibration of Agent-based Models (KDD Workshop 2024) \href{https://openreview.net/pdf?id=tgBrJUWon5}{[link here]}
    \item Chopra, Quera-bofarull and Zhang. differentiable agent-based modeling: systems, methods and applications (AAMAS 2024, Tutorial) \href{https://www.arnau.ai/diff_abms_tutorial/}{[link here]}
\end{itemize}

\begin{figure}
    \centering
    \begin{subfigure}[b]{0.43\textwidth}
        \centering
        \includegraphics[width=\textwidth]{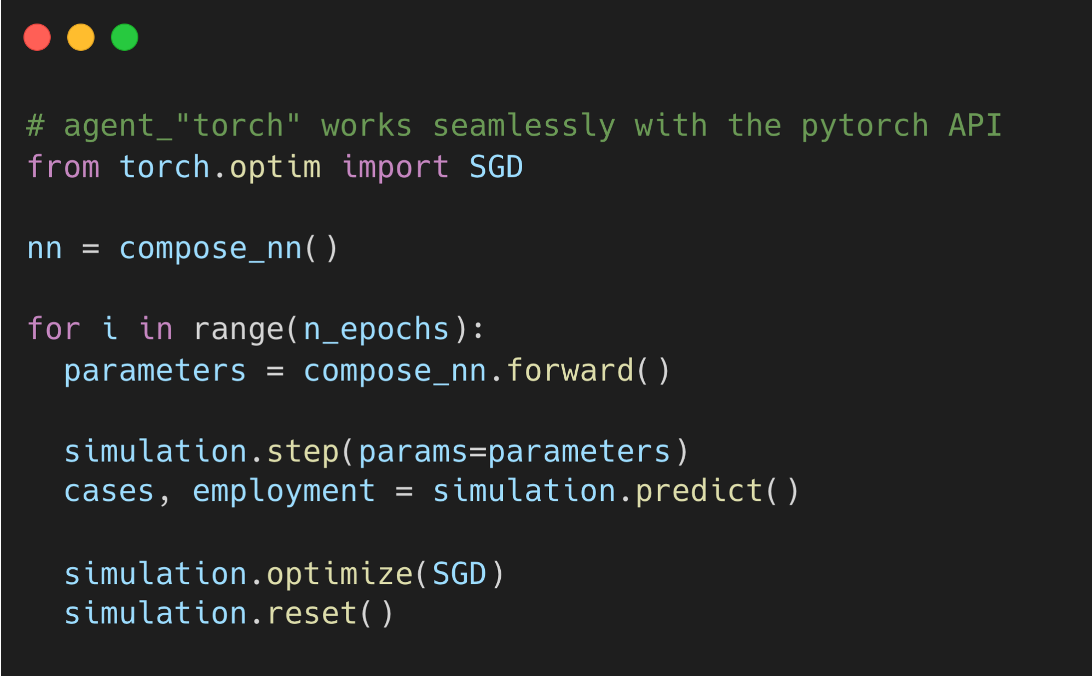}
        \caption{}
        \label{fig:diff_1}
    \end{subfigure}
    \hfill
    \begin{subfigure}[b]{0.48\textwidth}
        \centering
        \includegraphics[width=\textwidth]{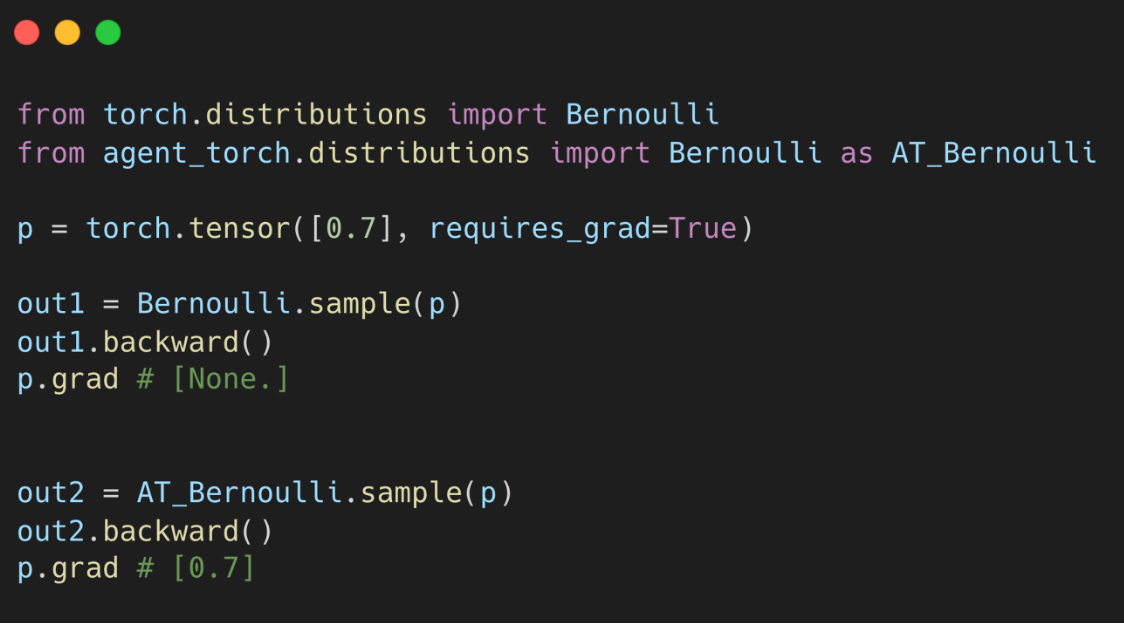}
        \caption{}
        \label{fig:diff_2}
    \end{subfigure}
    \caption{(a) Differentiable calibration pipeline showing how LPMs transform traditional black-box simulation into a differentiable computational object within neural parameter space, enabling gradient-based optimization. (b) Custom operators implementing differentiable stochastic mechanisms that maintain gradient flow through the simulation, preserving end-to-end differentiability for efficient calibration.}
\end{figure}

\begin{figure}[t!]
    \centering
    \includegraphics[width=0.95\linewidth]{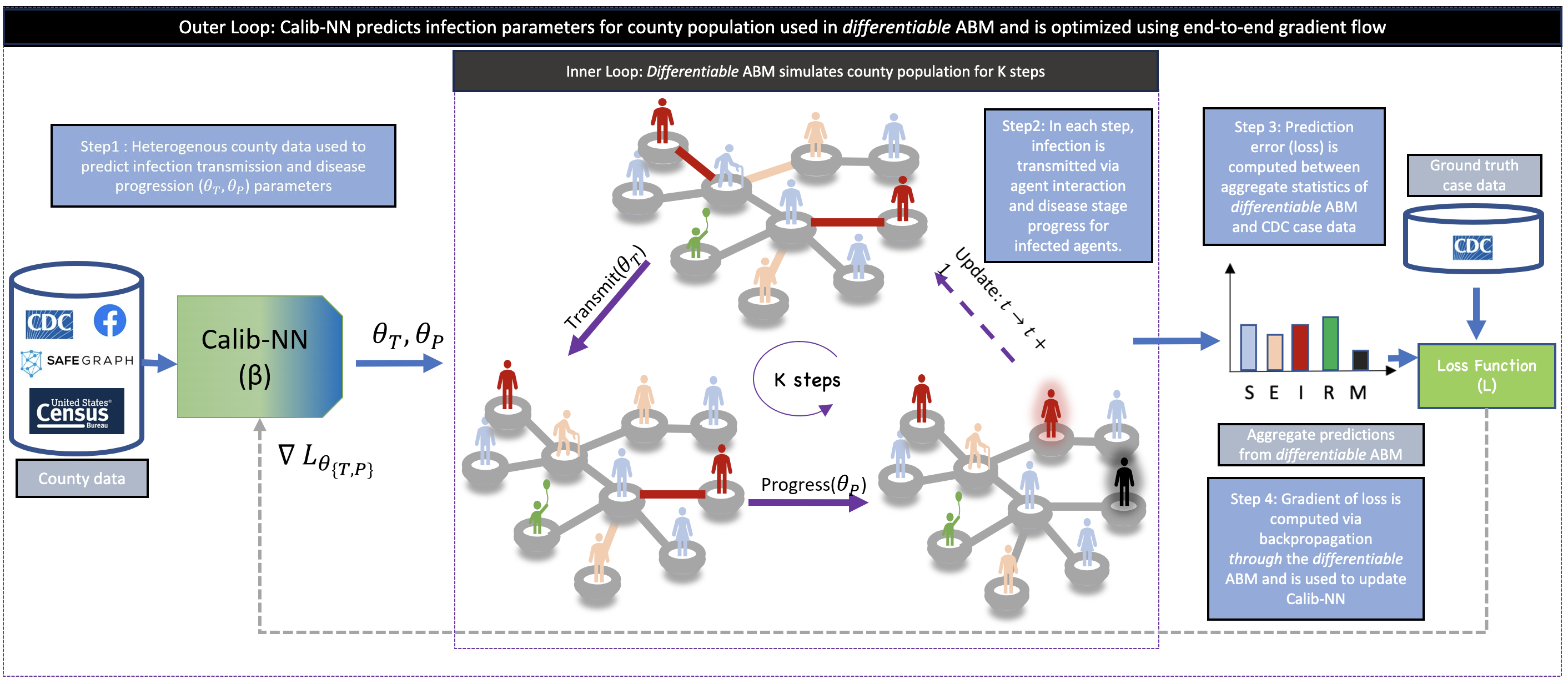}
    \caption{Calibration protocol for a differentiable ABM. Gradient-based learning helps leverage heterogeneous data sources, accelerates calibration time, improves robustness and was published at AAMAS 2023~\cite{chopraDifferentiableAgentbasedEpidemiology2023c}. The visualized protocol has four stages - i) heterogeneous data (CDC, census, behavioral, survey) is input to the ABM through DNN (calib-NN) to predict $\theta_T, \theta_P$), ii) ($\theta_T, \theta_P$) are used run $K$ forward steps of the fully-differentiable epidemiological model which simulates micro-level infection transmission ($\texttt{Transmit}$) and disease progression ($\texttt{Progress}$) over high-resolution individual contact networks. Disease statistics are aggregated ($\texttt{Aggregate}$) at end of $K$ steps to obtain the macro-level simulation output ($\hat{y}$). iii) Error between predicted $\hat{y}$ and real-world case statistics ($y$) is used to define a loss ($L(\hat{y}; \hat{y})$), iv) Gradient of this loss is computed by automatic differentiation through the differentiable ABM to update weights of calibNN}
    \label{fig:multimodal-calib}
\end{figure}

\subsection{Contribution 3: Rich Posterior Representation via Online Gradient-based Optimization}
Traditional ABM calibration approaches like Approximate Bayesian Computation (ABC) or Neural Likelihood Estimation (NLE) often use neural networks as surrogate models. However, these methods require offline sampling—running the simulator thousands of times to generate training data before optimization can begin. This offline approach suffers from poor sample efficiency and may not adequately explore the parameter space.

Our approach fundamentally differs by enabling online optimization directly on the simulation. We represent the variational family Q using deep neural networks:
$$\theta = \text{Sample}(\text{DNN}_Q; \phi)$$

We update these networks through online gradient-based learning that samples from the simulator during optimization. This delivers several advantages. First, each simulation run contributes directly to gradient updates, dramatically reducing the number of required simulation evaluations. Second, gradient information guides the sampling toward informative regions of parameter space, unlike offline approaches with fixed sampling schemes. Third, the neural parameterization captures multimodal, skewed, and complex dependencies that simple parametric distributions cannot represent.

For gradient-based optimization, we employ generalized variational inference to approximate the posterior distribution. The objective function is:
$$\mathcal{L}(\boldsymbol{\phi}) = \mathbb{E}{q{\boldsymbol{\phi}}} \left[ \ell(\mathbf{x}(\boldsymbol{\theta}), \mathbf{y})\right] + w; \mathrm{KL}(q_{\boldsymbol{\phi}}\mid\mid \pi(\boldsymbol{\theta}))$$
The gradient update decomposes elegantly as:
$$\nabla_{\boldsymbol{\phi}}\ell(\mathbf{x}(\boldsymbol{\theta}), \mathbf{y})= \nabla_{\boldsymbol{\phi}} \boldsymbol{\theta}{\boldsymbol{\phi}} \cdot \nabla{\boldsymbol{\theta}}\mathbf{x}$$

This decomposition enables efficient optimization: we estimate the left term via reverse-mode autodifferentiation and the right term using forward-mode autodifferentiation. This hybrid approach produces an efficient Jacobian computation that scales well even for large-scale simulations.

This online gradient-based approach eliminates the need for surrogate models altogether, directly optimizing the posterior representation while interacting with the simulation in real-time.

\subsection{Contribution 4: Data Assimilation for Collaborative Calibration}
Building upon the neural posterior representation, we extend our approach to incorporate heterogeneous data sources by conditioning the sampling process:
$$\theta = \text{Sample}(\text{DNN}_Q, \text{data}; \phi)$$

\begin{figure}[h!]
    \centering
    \includegraphics[width=0.7\linewidth]{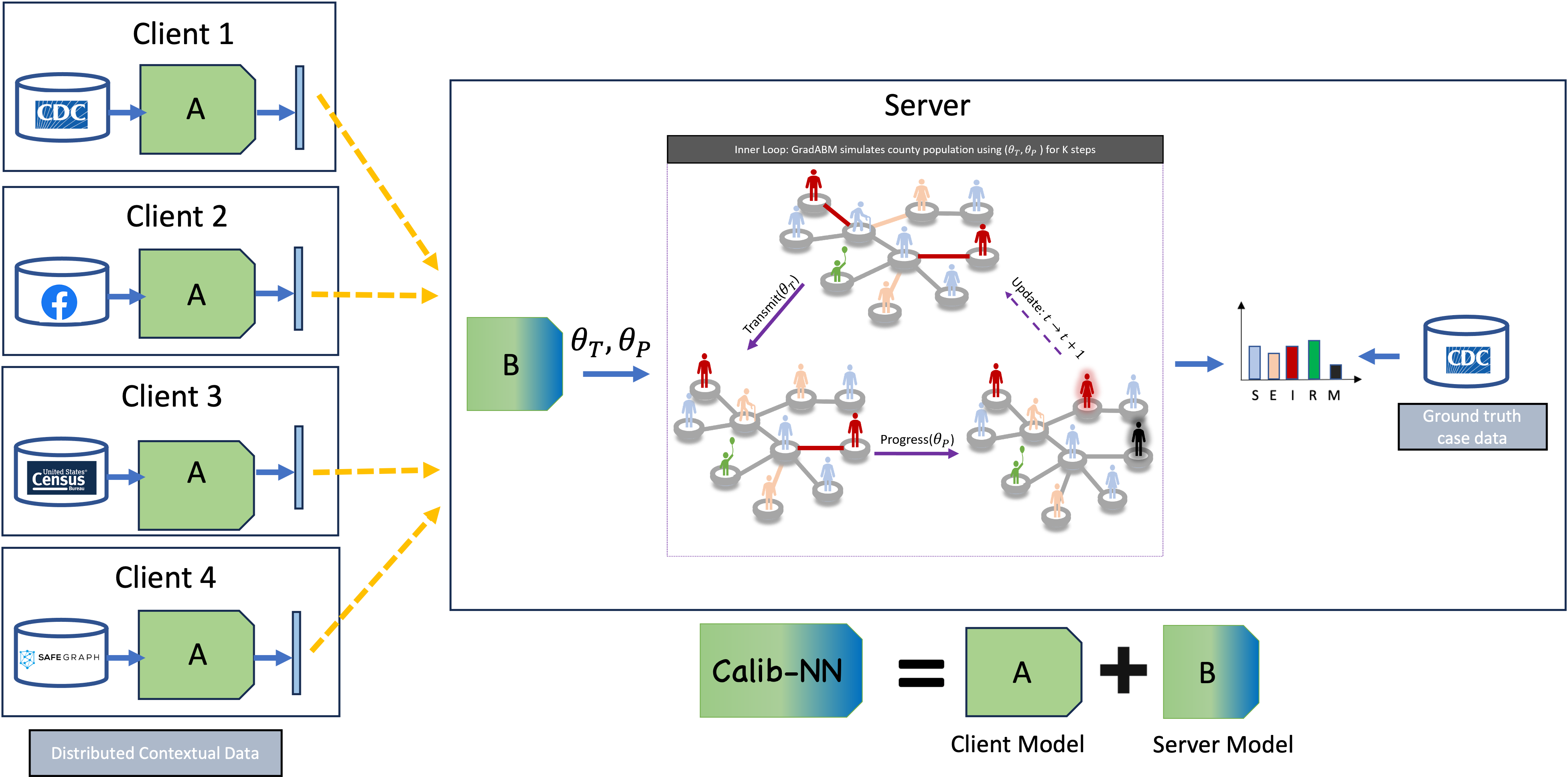}
    \caption{Distributed Calibration: We execute calibrate with contextual data distributed across multiple clients. To achieve this, we split CalibNN between the clients and server to ensure calibration on multi-modal data, without centralizing siloed information. Each client uses its local data to generate embeddings (\textcolor{orange}{in orange}) which are transmitted to server and used to predict structural parameters ($\theta_T, \theta_P$) to execute the simulation}
    \label{fig:dist-calib}
\end{figure}

This conditional generation framework provides several key capabilities. 
\begin{itemize}
    \item \textbf{Multi-task Calibration}: First, the neural network can ingest diverse data types including clinical reports, mobility patterns, and demographic information, extracting relevant features to inform parameter distributions. By conditioning on simulator-specific contexts, the same network architecture can be trained to calibrate multiple simulators simultaneously. This enables knowledge transfer across related domains and improves data efficiency. (figure~\ref{fig:multimodal-calib})
    \item \textbf{Distributed Calibration} Second, when sensitive data is siloed across institutions, we can optimize the network parameters through federated or split learning approaches. Each organization maintains local data while contributing to global parameter inference, addressing critical privacy concerns (figure~\ref{fig:dist-calib})
\end{itemize}

The data assimilation approach maintains all the benefits of rich posterior modeling from the previous section - the conditional distributions remain expressive and high-dimensional, avoiding restrictive assumptions about distribution shape or parameter independence. By allowing heterogeneous data sources to influence the parameter generation process directly, we enable more robust calibration that can adapt to complex, multi-modal evidence.

\begin{figure}[h!]
    \centering
    \includegraphics[width=0.75\linewidth]{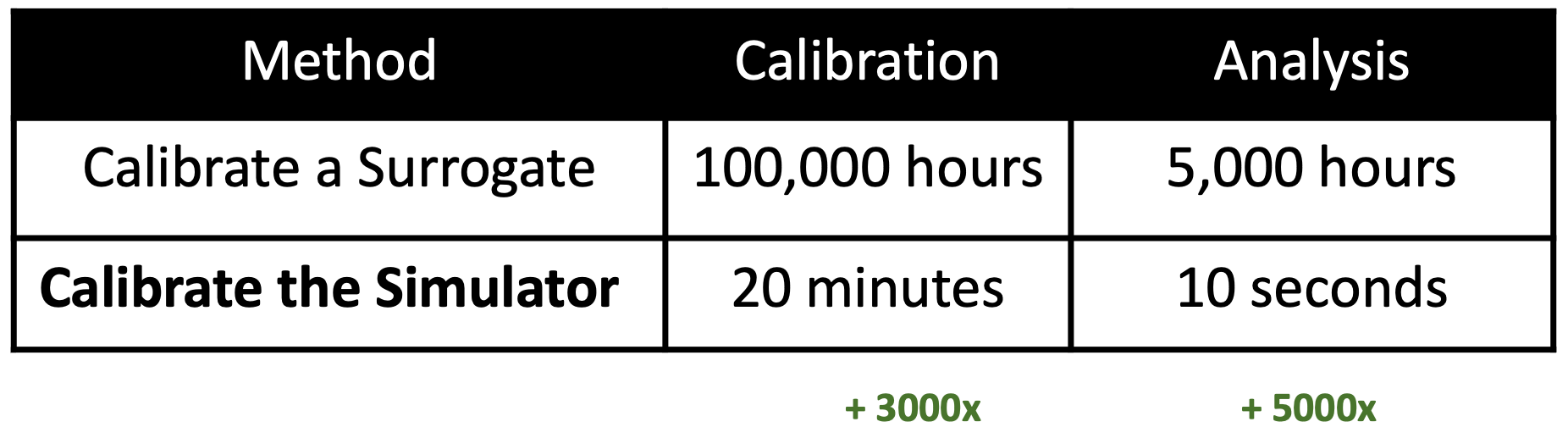}
    \caption{Comparative performance analysis between traditional surrogate-based calibration and direct simulation-based calibration, demonstrating how LPMs' differentiable approach provides substantial computational advantages: calibration time reduced from 100,000 hours to 20 minutes (3000x speedup) and analysis time reduced from 5,000 hours to 10 seconds (5000x speedup). This is possible since LPMs can: (i) compose multi-modal external data to explore calibration, (ii) execute sensitivity analysis using gradient trace in the computation graphs.}
    \label{fig:perform_calib}
\end{figure}

\subsection{Contribution 5: Zero-shot Sensitivity Analysis via AutoGrad}
A significant advantage of making the original simulation differentiable—rather than building a differentiable surrogate—is the ability to perform sensitivity analyses without additional simulation runs. After a single forward pass, we can compute exact partial derivatives of any output metric with respect to input parameters:
$$\frac{\partial \mathcal{M}}{\partial \boldsymbol{\theta}_i}$$

These derivatives are already computed and stored in the computational graph during the forward simulation, making them available "for free." This direct differentiation approach fundamentally differs from traditional sensitivity methods that require building a surrogate model from thousands of simulation samples, running additional simulations to estimate parameter sensitivity, and accepting approximation errors introduced by the surrogate.

By differentiating the actual simulation rather than an approximation, we eliminate surrogate modeling errors while dramatically reducing computational costs. This zero-shot capability enables simultaneous sensitivity evaluation for all parameters, precise quantification of parameter influence without approximation errors, sensitivity information derived from the actual simulation mechanics rather than an emulator, and no additional simulation runs required beyond the initial calibration.


This approach represents a fundamental shift in how sensitivity analysis is conducted for complex simulations, making comprehensive validation practical even for large-scale models with millions of agents. The ability to quickly identify influential parameters and quantify their effects supports both model refinement and transparent communication of results to decision-makers.

\section{Challenge 3: Simulation-Reality Gap}
The scalable and differentiable simulation approaches we've developed in previous sections enable modeling millions of synthetic agents with sophisticated behaviors. However, these innovations in computational efficiency are of limited utility if the quality of underlying data is poor. Currently, agent-based simulations rely primarily on aggregate census and mobility data that has been anonymized~\cite{chapuisGenerationSyntheticPopulations2022a} or made differentially private~\cite{aktay2020google}, significantly limiting model expressiveness and predictive power.

The utility of simulations is hence limited in informing real-world interventions due to a persistent gap between simulation insights and operational implementation. While our models can simulate complex behaviors at scale, they often rely on synthetic or highly anonymized data that insufficiently captures the granularity of real-world interactions.

During COVID-19 pandemic, health authorities attempted to close this sim2real gap by capturing real-time individual data via contact tracing apps. However, this approach faced two key limitations:
\begin{itemize}
    \item \textbf{Privacy concerns limit data quality}: Attempts to centralize data from contact tracing apps led to significant privacy breaches - with sensitive mobility or health data of individuals was leaked and misused for surveillance ~\cite{IndonesiaProbesSuspected2021, kelleyIllinoisBoughtInvasive2021, coxTMobilePutMy2019}. These incidents reduced adoption rates and led to stricter privacy regulations, resulting in highly anonymized or synthetic data that lacks the granularity needed for accurate simulations.
    \item \textbf{Timeliness challenges prevent proactive modeling}: By the time data is collected, anonymized, and integrated into centralized simulations, the world has changed. Contact tracing apps could notify users about past exposures, but cannot support forward-looking simulation tasks like exposure management, intervention evaluation, or risk minimization—activities that require real-time data integration.
\end{itemize}

This creates a fundamental tension: bringing data to simulations introduces privacy vulnerabilities and time delays that significantly diminish utility, yet simulations require detailed, current data to provide actionable insights. The tension between privacy, timeliness and effectiveness severely limited adoption-with contact tracing apps reaching only 21\% of the US population at peak usage.

Consider the COVID-19 response: contact tracing apps could notify users of exposure but lacked the simulation capabilities to answer critical questions like "How can I minimize future risk?" or "Which interventions would most effectively reduce community transmission?" By the time contact data reached central modeling teams for analysis, the dynamics had already shifted, making insights outdated before they could be implemented. These challenges reflect a deeper issue—our inability to create a secure bridge between simulation models and the real-world agents they represent. 

\textbf{Key Insight}: The ideal solution is not to bring data to simulations, but rather to bring simulations to data—enabling secure, decentralized computation directly where data resides while maintaining individual privacy. LPMs introduce a dual notion of an agent - an entity that can exist in both synthetic and physical environments. The key innovation is extending differentiable simulation capabilities to distributed agents while preserving individual privacy, allowing simulations to operate on fresh, granular data without introducing privacy risks. We achieve this through additive secret sharing which allows secure computation of simulation outputs and gradients.

This is represented in the following publications:
\begin{itemize}
    \item Chopra et al. Private Agent-based Modeling. (AAMAS 2024, Oral) \href{https://arxiv.org/abs/2404.12983}{[link here]}
\end{itemize}



\subsection{Contribution 6: Decentralized Modeling with Physical Agents}
We introduce novel protocols that enable secure, decentralized simulation, calibration, and analysis of agent-based models. Rather than centralizing sensitive data, our approach leverages secure multi-party computation (MPC) to perform model operations directly where data resides—on individual devices or institutional systems—while maintaining privacy guarantees. 

This conceptualization transforms population modeling from a retrospective analysis tool into an integrated component of adaptive response systems. This addresses both the privacy and timeliness challenges of traditional approaches, enabling simulations to operate on fresh, granular data without introducing privacy risks. With LPMs, we can now - estimate personalized risk based on current contact patterns, calibrate disease models using privacy-preserved individual-level data, evaluate policy interventions without compromising individual behavioral response.

Specifically, in LPMs, FLAME protocols preserve the same mathematical properties in decentralized computation as a centralized simulations without requiring agents to reveal their private attributes or interactions. This enables LPMs to extend differentiable simulation capabilities to distributed agents while preserving individual privacy and simulation accuracy.

\textbf{Formal Notation}: Consider the base formulation in Equation~\ref{eq:agent_update} and two critical privacy-sensitive elements emerge:
\begin{enumerate}
    \item Agent state variables $s_i(t)$, which may include personal attributes like health status, demographic information, or location
    \item Neighborhood interaction data $\bigoplus_{j\in N_i(t)} m_{ij}(t)$, which reveals both social connections and transmission pathways
\end{enumerate}

In the context of an epidemiological model, imagine the transmission protocol of a contact-based disease simulation where an agent's infection probability depends on their interactions with potentially infectious neighbors. Specifically, agent $i$ updates its state following Equation~\ref{eq:sim_eqn} with message function defined as:
\begin{equation}
M_{ij}(t) = I_j(t)
\end{equation}

where $I_j(t)$ is the infected status of neighbor $j$ (0 or 1). The probability of disease risk for agent $i$ is then:

\begin{equation}
p^{(i)}_{\text{inf}}(t) = 1 - \exp\left(-\frac{\beta S_i\Delta t}{n_i}\sum_{j\in N(i)}I_j(t)\right)
\end{equation}

where $N(i)$ is the set of neighbors of agent $i$, $S_i$ is the susceptibility of agent $i$, $n_i = \#N(i)$ is the total number of neighbors, $\Delta t$ is the duration of the time-step, and $\beta$ is a structural parameter called the effective contact rate.

Computing this probability directly would reveal both the infection status of each neighbor and the complete neighborhood structure—creating significant privacy risks. Centralized simulations typically handle this by using coarse, noisy approximations of mobility and contact patterns, which significantly limits model fidelity.

\textbf{Additive Secret Sharing for Simulation Outputs and Gradients}: We overcome this challenge through additive secret sharing protocols that enable computation over sensitive data without revealing individual values. The key insight is to divide a secret input into multiple shares such that the secret can be reconstructed only when a sufficient number of shares are combined.

Consider $N$ agents holding private values $s_i$ (such as infection status). We want to compute the sum $\sum_i s_i$ without any agent $j$ acquiring knowledge about $s_{k\neq j}$. Each agent $i$ samples $N-1$ random numbers, $r_{ij} \sim U\{0, n-1\}$, such that the input is divided into $N$ shares, $s_{ij}$ defined by:

\begin{equation}
s_i = \sum_{j=1}^{N}s_{ij} \pmod{n} = \sum_{j=1}^{N-1}r_{ij} + \left(s_i - \sum_{j=1}^{N-1}r_{ij}\right) \pmod{n}
\end{equation}

Each agent distributes these shares (keeping one) and locally performs:
\begin{equation}
\sigma_k = \sum_{i=1}^{N}s_{ik} \pmod{n}
\end{equation}

\begin{figure}
        \centering
    \begin{subfigure}[b]{0.45\textwidth}
        \centering
        \includegraphics[width=\textwidth]{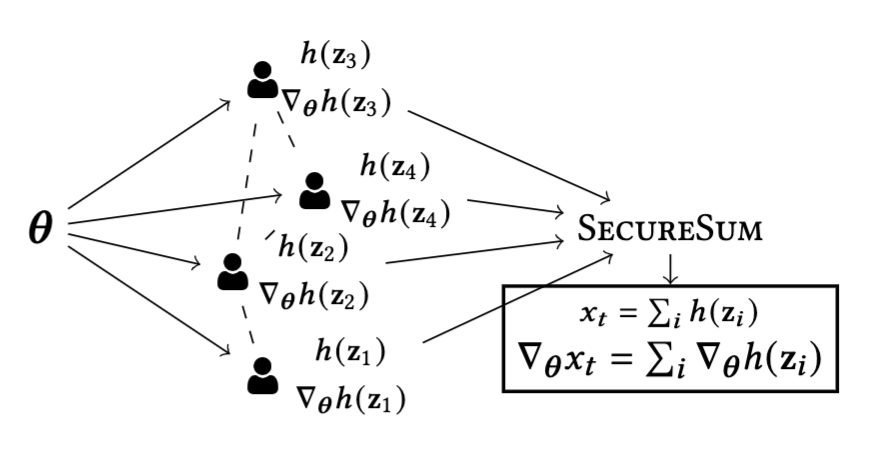}
        \caption{}
        \label{fig:flame_1}
    \end{subfigure}
    \hfill
    \begin{subfigure}[b]{0.45\textwidth}
        \centering
        \includegraphics[width=\textwidth]{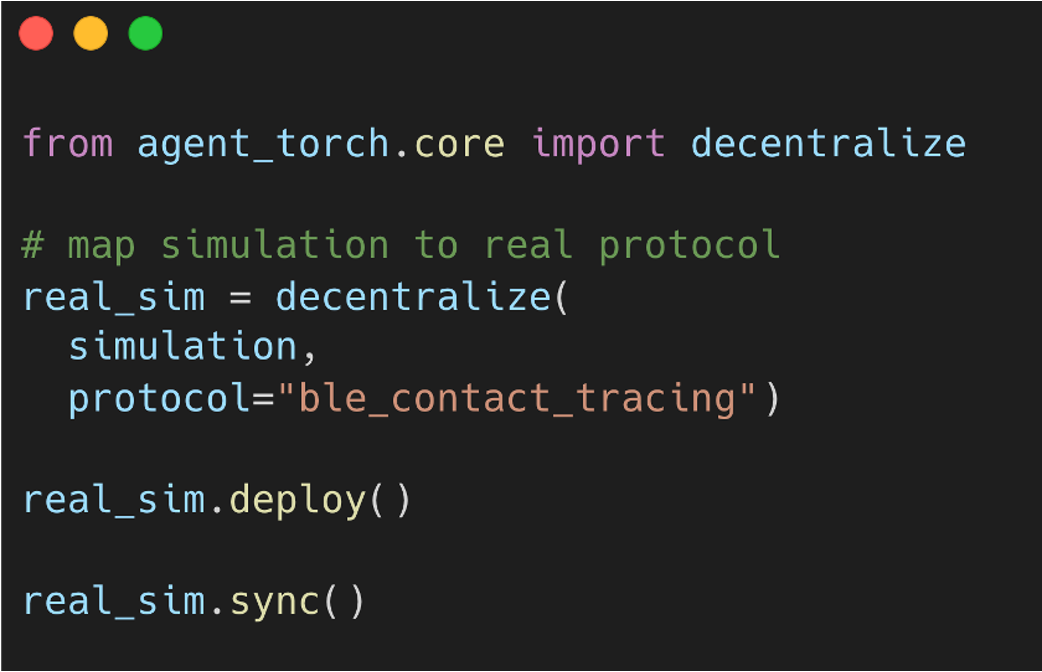}
        \caption{}
        \label{fig:flame_2}
    \end{subfigure}
    \caption{(a) Secure computation architecture showing how LPMs enable secure aggregation of neighborhood messages and gradients through additive secret sharing, maintaining privacy while preserving simulation accuracy. (b) AgentTorch API for decentralized simulation deployment, demonstrating how simulation logic can be mapped to distributed physical agent nodes while maintaining privacy guarantees.}
    \label{fig:enter-label}
\end{figure}

The final result can be securely reconstructed as $S = \sum_k \sigma_k \pmod{n}$.

Applied to our disease transmission example, this enables each agent to compute their infection probability without learning the actual infection status of any specific neighbor. 

Similarly, for calibration and analysis, each agent can compute the gradient $\nabla_{\boldsymbol{\theta}} \mathbf{x}$, where $\mathbf x$ is the number of daily infections and $\theta = \beta$. We note that this gradient can be approximated by the gradient of the average number of new infections with respect to $\beta$,
\begin{equation}
    \frac{\partial x_t}{\partial\beta} \approx \frac{\partial \; \mathbb{E}[\Delta I(t)]}{\partial \beta} = \sum_{i=1}^N \chi_i(t) \exp(-\chi_i(t) / \beta),
\end{equation}
where
\begin{equation}
    \chi_i(t) = \exp\left(-\frac{\beta\, S_i\Delta_t}{n_i} \sum_{j \in \mathcal N(i)} I_j(t) \right).
\end{equation}

Finally, these gradients $\nabla_\theta h(s_i(t))$ computed locally, and these can be securely aggregated through:

\begin{equation}
\nabla_\theta x_t = \bigoplus_{i\in A}\nabla_\theta(h(z_i(t)))
\end{equation}

This approach maintains the mathematical properties of a centralized simulation while preserving privacy, enabling sophisticated modeling capabilities without compromising security.

\section{Open Problems}
Large Population Models is an emerging research direction with several problems at the intersection of agent-based modeling, decentralized learning and machine learning. I provide a list of ideas with contextual motivation following from the tasks above:

\subsection{Scalable Agent-based Simulations}
This pillar focuses on efficiently simulating millions of interacting agents.

\paragraph{Problem 1: Group Archetypes}
The individual-based archetype approach fails to capture the interdependent nature of decision-making within social units. During the COVID-19 pandemic, we observed that household-level decisions regarding stimulus spending, isolation behaviors, and vaccination timing emerged from complex intra-group negotiations rather than independent individual choices. Current LPM implementations sample agent decisions independently, which misses critical correlation structures and leads to unrealistic emergent behaviors, particularly when modeling economic interventions like stimulus payments that target household units rather than individuals.

\paragraph{Problem 2: Formal Verification of Simulations}
The FLAME domain-specific language currently relies on empirical validation rather than formal guarantees, creating potential risks when composing heterogeneous modeling paradigms (discrete events, continuous dynamics, stochastic processes) in high-stakes policy simulations. During the NYC pandemic modeling, we observed numerical instabilities and unexpected emergent behaviors when combining economic, mobility, and disease transmission substeps, raising questions about whether these compositions preserved critical invariants across simulation components. The fundamental challenge remains: what spatial resolution to simulate each substep at?


\subsection{Differentiable Agent-based Simulations}
This pillar focuses on efficiently computing simulation gradients over synthetic agents.

\paragraph{Problem 3: Gradient Estimators for Discrete Randomness} 
LPMs frequently involve discrete stochastic decisions (e.g., whether to isolate after exposure, or whether to adopt a contact-tracing app), yet current gradient-based optimization approaches struggle with these non-differentiable components. In our NYC case study, calibration of behavioral parameters governing binary decisions introduced significant variance in gradient estimates, leading to unstable optimization and requiring excessive sampling to achieve reliable results. This fundamentally limits our ability to efficiently calibrate LPMs with realistic discrete choice components.

\paragraph{Problem 4: Simulator as Predictor}
Despite LPMs' mechanistic interpretability advantages, their adoption in decision-critical contexts is hindered by the perception that black-box time-series forecasting models achieve superior predictive accuracy. During COVID-19 response, NYC officials often had to choose between interpretable but potentially less accurate LPMs versus accurate but opaque statistical forecasts. This created a false dichotomy between understanding and accuracy that undermined trust in simulation-based policy recommendations.


\subsection{Decentralized Agent-based Simulations}
This pillar focuses on securely deploying simulations over decentralized networks of physical agents.

\paragraph{Problem 5: The Cold Start Challenge}
Decentralized agent networks suffer from adoption challenges due to limited initial utility. During COVID-19, contact tracing apps reached only 21\% adoption in the US, creating a cold start problem where the app's usefulness depended on network effects that couldn't materialize without sufficient initial adoption. This fundamental challenge—providing value before reaching critical mass—limits the practical deployment of decentralized LPMs regardless of their theoretical capabilities.

\paragraph{Problem 6: Orchestrating Agent Networks} Complex socio-technical systems often face incentive misalignment between individual agent objectives and global welfare. During NYC's pandemic response, we observed this misalignment when rational individual behaviors (e.g., mobility decisions optimizing for personal utility) led to suboptimal collective outcomes (e.g., increased disease transmission). Current approaches either assume unrealistic compliance with centralized directives or rely on simplified rational agent models that fail to capture the nuanced, heterogeneous behaviors observed in real populations. LPMs offer the potential to discover robust incentive mechanisms that align decentralized decision-making with social welfare objectives, without requiring centralized control or data sharing.

\section{Conclusion}
Large Population Models (LPMs) represent a significant advancement in our ability to understand and address challenges that emerge from the interactions of millions of individuals. By overcoming the limitations of traditional agent-based models, LPMs enable more accurate predictions, more efficient policy evaluation, and more seamless integration with real-world systems.

The three key innovations presented in this paper—compositional design with tensorized execution, differentiable specification, and decentralized computation—work together synergistically to transform population-scale modeling. Compositional design allows us to efficiently simulate millions of agents with sophisticated behaviors on commodity hardware. Differentiability enables gradient-based calibration and learning from heterogeneous data streams. Decentralized protocols bridge the gap between simulated and physical environments while protecting individual privacy. This opens new possibilities for addressing challenges from pandemic response to climate adaptation— characterized by complex feedback loops between individual behavior and population-scale dynamics.


The COVID-19 case study illustrates how LPMs can provide actionable insights in crisis scenarios by capturing the intricate interplay between disease transmission, economic impacts, and behavioral adaptation. These capabilities extend beyond epidemiology to any domain where individual decisions aggregate into collective outcomes that, in turn, reshape individual incentives.

As digital technologies increasingly mediate social interactions and generate unprecedented data about human behavior, LPMs provide a framework for responsibly leveraging this information to understand and improve social systems. By combining computational efficiency, mathematical rigor, and privacy-preserving protocols, LPMs offer a path toward modeling complex social dynamics at true population scale while respecting individual autonomy and privacy.

While current AI advances primarily focus on creating sophisticated individual agents, LPMs highlight the importance of understanding collective intelligence and emergent phenomena. As we continue to develop more powerful computational tools, LPMs serve as a reminder that many of our most pressing challenges require not just modeling individual cognition, but the complex web of interactions through which individual behaviors become social outcomes.

\bibliography{lpm_papers}
\bibliographystyle{plain}






\newpage

\end{document}